# $CuIr_2Te_4$: A Quasi-Two-Dimensional Ternary Telluride Chalcogenide Superconductor

Dong Yan[1#], Yijie Zeng[2#], Guohua Wang[3], Yiyuan Liu[4], Junjie Yin[2], Tay-Rong Chang[5,6], Hsin Lin[7], Meng Wang[2], Jie Ma[3], Shuang Jia[4], Dao-Xin Yao[2*], Huixia Luo[1*]

[1] *School of Material Science and Engineering, Key Lab of Polymer Composite & Functional Materials, Sun Yat-Sen University, No. 135, Xingang Xi Road, Guangzhou, 510275, P. R. China*

[2] *School of Physics, State Key Laboratory of Optoelectronic Materials and Technologies, Sun Yat-Sen University, No. 135, Xingang Xi Road, Guangzhou, 510275, P. R. China*

[3] *Department of Physics and Astronomy, Shanghai Jiao Tong University, Shanghai 2 00240, China*

[4] *Department of Physics, Peking University, Beijing, 100871, China*

[5] *Department of Physics National Cheng Kung University Tainan 701, Taiwan*

[6] *Center for Quantum Frontiers of Research and Technology (QFort), Tainan 701, Taiwan*

[7] *Institute of Physics, Academia Sinica, Taipei 11529, Taiwan*

[#] These two authors made equal contributions

[*]Corresponding author:

H.X. Luo, E-mail address: luohx7@mail.sysu.edu.cn;

D. X. Yao, E-mail address: yaodaox@mail.sysu.edu.cn

**ABSTRACT:** Here we report the first observation of superconductivity in the $AB_2X_4$-type ternary telluride $CuIr_2Te_4$, which is synthesized by a solid-state method in an evacuated quartz jacket. It adopts a disordered trigonal structure with space group $P\bar{3}m1$ (No. 164), which embodies a two-dimensional (2D) $IrTe_2$ layers and intercalated by Cu between the layers. We use a combination of experimental and first principles calculation analysis to look insight into the structural and physical properties. $CuIr_2Te_4$ consistently exhibited a bulk superconductivity transition at 2.5 K in electrical resistivity, magnetic susceptibility and specific heat measurements. Resistivity and magnetization measurements suggest a charge density wave transition ($T_{CDW} \approx 250$ K) coexists in $CuIr_2Te_4$, which further signify the coexistence of superconductivity and CDW in the ternary telluride chalcogenide $CuIr_2Te_4$. Our discovery of the new CDW-bearing superconductor $CuIr_2Te_4$ opens a door for experimental and theoretical studies of the interplay between CDW and superconductivity quantum state in the condensed matter field.

## INTRODUCTION

The $AB_2X_4$ chalcogenide system ($A$, $B$ = transition metal ions; $X$ = S, Se, Te) have attracted continuous research interests due to their rich structure types and abundant physical properties.[1-5] The ternary and pseudo-ternary $AB_2X_4$ chalcogenides generally adopt the cubic spinel structure (space group $Fd\bar{3}m$), monoclinic defect NiAs structure (space group $I2/m$) or the disordered trigonal structure (space group $P\bar{3}m1$).[6-8] Many $AB_2X_4$-type ternary and pseudo-ternary sulfo- and seleno-spinels have been discovered in past decades, but there are few examples of superconducting spinels.[9]

$CuV_2S_4$ sulfo-spinel is one of the earliest three-dimensional $AB_2X_4$-type materials known to superconduct ($T_c$ = 4.45-3.20 K) and strikingly exhibits three charge-density wave (CDW) states ($T_{CDW1}$ = 55 K, $T_{CDW2}$ = 75 K, $T_{CDW3}$ = 90 K).[10-11] The famous $CuRh_2S_4$ and $CuRh_2Se_4$ with the typical cubic spinel structure are also well known superconductors with $T_c$ = 4.35 and 3.50 K, respectively.[12-16] Yet, another popular ternary metal chalcogenide spinel compound, $CuIr_2S_4$, doesn't superconduct but exhibits a quite odd metal-insulator (M-I) transition at T $\approx$ 230 K, which is followed by a complex structural transition that simultaneously produces both charge ordering and metal-metal pairing.[17] Interesting, the M-I transition can be suppressed and domelike superconductivity phase diagram ($T_c(x)$) was discovered by Zn substitution for Cu in the $Cu_{1-x}Zn_xIr_2S_4$ solid solution, with a maximum $T_c$ of 3.4 K near $x$ = 0.3.[18] The other prototypical ternary metal seleno-chalcogenide $CuIr_2Se_4$ with the same spinel structure as $CuIr_2S_4$, remains metallic behavior from room temperature down to 0.5 K. Neither M-I or nor superconducting transitions are observed at ambient pressure, but a M-I transition is found above 2.8 GPa.[19-21] Surprisingly, a superconductivity was observed in $Cu(Ir_{1-x}Pt_x)_2Se_4$ ($0.1 \leq x \leq 0.35$) with a maximum $T_c$ = 1.76 K near $x$ = 0.2 with Pt substitution for Ir in the $CuIr_{2-x}Pt_xSe_4$ solid solution.[22] Despite hundreds of $AB_2X_4$-type materials have been discovered, the superconductors found in this family are only the aforementioned $CuV_2S_4$, $CuRh_2S_4$, $CuRh_2Se_4$, electron-doped $CuIr_2S_4$ and electron-doped $CuIr_2Se_4$ sulpho- or seleno- spinels. All of the above mentioned superconducting $AB_2X_4$-type compounds have the normal spinel structure where the Cu ions occupy the A tetrahedral sites and the transition metal ions (V, Rh, Ir) occupy the B octahedral sites. To the best of our knowledge, no superconductivity has been observed in $AB_2X_4$-type telluro-compounds or $AB_2X_4$-type sulpho- and seleno-compounds with defect structures related to NiAs so far.

Currently, it is commonly approved that low dimensionality leads to particular electronic structures and allows relatively strong fluctuations, which may improve superconductivity. Although CDW sometimes contests with superconductivity, mainly in the quasi-one-dimensional case.[23] From a crystal-chemistry perspective, low-dimensional structures easily appear in inorganic metal compound with relatively high polarizability and low electronegativity anions. For instance, by comparison with metal oxides, metal chalcogenides have a tendency to crystallize in layered structures,[24-25] which leads to rich and interesting physical phenomena such as CDW and superconductivity.[23,26] Thus, it is wise to search for new superconductors in the $AB_2X_4$-type chalcogenides with low-dimensional crystal structures. On the other hand, among chalcogenides, the crystal structures electronic structures and physical properties of tellurides are often different from sulfides and selenides, since Te has the largest ionic radius among S, Se and Te. About two decades ago, Nagata et al. observed anomalies in resistivity and magnetic susceptibility in $CuIr_2Te_4$, probably resulting from the formation of CDW.[27] However, its low-temperature properties have not been characterized. Unlike $CuIr_2S_4$ and $CuIr_2Se_4$ spinels, $CuIr_2Te_4$ crystallizes in a two-dimensional (2D) disordered trigonal structure with the space group *P*-3*m*1 at room temperature, which embodies a complicated 2D $IrTe_2$ layers and intercalated by Cu between the layers (**Figure 1a**). On first sight though no superconducting transition was observed from ambient temperature down to 4.2 K, it seems to be an ordinary metal with a CDW transition around 250 K. Only limited reports have been given by earlier researchers, and no complete studies and analysis related with superconductivity are yet explored. However, it is well-known that many of the two-dimensional dimensional transition metal dichalcogenides (TMDCs) exhibit CDW and some of them displays the competition/coexistence between CDW and superconductivity. Inspired by the 2D TMDCs, we quested for the coexistence of CDW and superconductivity in the 2D ternary telluride chalcogenide $CuIr_2Te_4$ of $AB_2X_4$ type.

In the present contribution, we report the first observation of superconductivity in $AB_2X_4$-type ternary telluride $CuIr_2Te_4$, which was successfully synthesized trough solid-state reaction method. The structural properties of the $AB_2X_4$-type telluro-compound $CuIr_2Te_4$ was characterized via X-ray diffraction (XRD), energy dispersive X-ray spectrometer (EDXS) and high-resolution transmission electron microscopy (HRTEM). Meanwhile, the superconductivity properties were systematically studied *via* combining with experimental and first principles calculation analysis. The results of magnetization, resistivity, and heat capacity measurements demonstrate that $CuIr_2Te_4$ is a quasi-two-dimensional ternary telluride chalcogenide superconductor with $T_c \approx 2.5$ K and suggest that a CDW state emerges below around 250 K. With the discovery of this CDW-bearing superconductor $CuIr_2Te_4$, there is large room for further explorations of the interplay between CDW and superconductivity in the $AB_2X_4$ system.

## EXPERIMENTAL

### Method and materials

Polycrystalline $CuIr_2Te_4$ was prepared by a conventional solid-state reaction method. Mixtures of high-purity fine powders of Cu (99.5%), Ir (99.95%) and Te (99.999%) in the appropriate stoichiometric ratios were thoroughly ground, pelletized and heated in a sealed quartz tube at 850 °C for 96 h. Subsequently, the as-prepared powders were reground, pelletized and sintered at 900 °C for 96 h.

We also adopted the powder X-ray diffraction (PXRD, Bruker D8 focus, Cu Kα radiation, graphite diffracted beam monochromator) to characterize the samples structurally. The unit cell parameters were determined by profile fitting the powder diffraction data with FULLPROF diffraction suite with Thompson-Cox-Hastings pseudo-Voigt peak shapes.[28] A Quantum Design Physical Property Measurement System (PPMS) was used to take the measurements of the temperature dependence of electrical resistivity and heat capacity from 2 K to 300 K. No materials' air-sensitivity was found during the measurements. $T_c$ determined from the susceptibility data were estimated conservatively: $T_c$ was taken as the intersection of the extrapolations of the steepest slope of the susceptibility in the superconducting transition region and the normal state susceptibility; $T_c$ was determined from the midpoint of the resistivity $\rho$(T) transitions for the resistivity data; for the specific heat data, we employed equal area construction method to obtain the critical temperature.

The microscopic morphology and the composition of $CuIr_2Te_4$ sample were investigated by scanning electron microscopy (SEM, Quanta 400F, Oxford) and energy dispersive X-ray spectroscopy (EDXS), respectively. And high-resolution transmission electron microscopy (HRTEM, FEI Tecnai G2 F20 operated on 200 kV) was used to study the phase structure of $CuIr_2Te_4$.

**Calculational Details**

The first principles calculation was carried out by using the projector augmented wave method [29] implemented in the VASP package.[30] The generalized gradient approximation suggested by Perdew, Burke and Ernzerhof [31] was used for the exchange-correlation interaction. The cutoff energy was set to 900 eV and a Monkhorst-Pack grid [32] of 15 × 15 × 11 was used for *k*-point sampling. The spin-orbit coupling was considered in the band structure calculation.

Phonon dispersion was calculated by finite displacement method as implemented in Phonopy[33], with a supercell of 3 × 3 × 2. Since the occupation of Cu is 0.5 at 1*b* site (see **Table 1**), we built the primitive unit cell by doubling the unit cell of $P\bar{3}m1$ along *c* and preserved only one Cu atom. Physically this corresponds to the case that Cu was intercalated by every two $IrTe_2$ layers. We will see this is a reasonable model as it gives results in agreement with experiment semi-quantitatively.

**RESULTS AND DISCUSSION**

**Figure 1** shows powder XRD characterization for polycrystalline $CuIr_2Te_4$. Except for tiny impurities of Ir, all the reflection peaks can be indexed to the $CuIr_2Te_4$ compound, which adapts to the disordered trigonal structure with the space group $P\bar{3}m1$ at room temperature. $CuIr_2Te_4$ exhibits a two-dimensional disordered trigonal structure, which embodies two-dimensional (2D) $IrTe_2$ layers intercalated by Cu between the layers. (**Figure 1a**). The lattice parameters are obtained to be *a* = b = 3.9397(3) Å and *c* = 5.3964(6) Å (**Table 1**). **Figure 1d** is the scan electron microscope (SEM) images of the $CuIr_2Te_4$ polycrystalline powder sample, which signify $CuIr_2Te_4$ has a layer structure. The HRTEM image (**Figure 1b**) of $CuIr_2Te_4$ polycrystalline powder shows clear lattice distance of 0.510 nm. Fast Fourier transform images (FFT) are shown in the upper right corner of the image. **Figure 1c** shows the [001] zone axis pattern (ZAP) of $CuIr_2Te_4$. It indicates that $CuIr_2Te_4$ has a trigonal structure; which is consistent with the XRD data previously mentioned. The EDXS spectrum (**Figure 1e**) shows three elements of Cu, Ir and Te in the as-prepared $CuIr_2Te_4$ polycrystalline powder sample. And the element ratio of Cu, Ir and Te is close to 1: 1.9: 4, indicating that the $CuIr_2Te_4$ polycrystalline powder compound is close to stoichiometric ratio.

The cooling and warming temperature dependence of electrical resistivity ($\rho$(T)) for polycrystalline $CuIr_2Te_4$ under zero magnetic field is shown in the main panel of **Figure 2a**. The samples show a metallic temperature dependence ($d\rho/dT$) in the region of 2.7 to 300 K. The up-left corner inset of **Figure 2a** presents the $d\rho/dT$ *vs.* $T$ curve at low temperature, which further confirms the $T_c$ is around 2.5 K. A sharp drop of $\rho$(T) obviously takes place below 2.7 K (bottom right corner inset of **Figure 2a**), which represents the occurrence of superconductivity and demonstrates the details of the superconducting transition. Superconductivity is also confirmed by the zero-field-cooled (ZFC) and field-cooled (FC) *dc* magnetic susceptibility measurements (**Figure 2b**), showing a strong diamagnetism below 2.5 K. The superconducting volume fraction can be estimated approximately to be 96.4 %, which reveals the high purity of the $CuIr_2Te_4$ sample. The inset of **Figure 2b** shows the $d\chi/dT$ *vs* $T$ curve at low temperature, revealing a sharp peak at the superconducting transition temperature. As shown in **Figure 2c**, the susceptibility exhibits an anomaly around 250 K with the hysteresis, which indicate the CDW transition at around 250 K and is in consistent with the previous report. [27] Inset of **Figure 2c** shows the measured temperature range from 150 to 300 K of electrical resistivity ($\rho$(T)) at applied field in the range from 0 to 9 T with cooling and heating. Resistivity measurements all exhibited a distinct phase transition associated with the formation of CDW around 225 - 300 K, which was not affected by the applied magnetic fields. The above results reveal that $CuIr_2Te_4$ is a new CDW-bearing superconductor.

Temperature-dependent measurements of the magnetization under incremental magnetic field M(H) were applied to determine the upper critical field $\mu_0H_{c1}(0)$. **Figure 2d** shows how the $\mu_0H_{c1}(0)$ for $CuIr_2Te_4$ was determined. First, applied field magnetization measurements M(H) were performed at 1.8, 2.0, and 2.2 K to calculate the demagnetization factor (N). With the hypothesis that the beginning linear response to the magnetic field is perfectly diamagnetic (dM/dH = − 1/4 $\pi$) for this bulk superconductor, we obtained the values of demagnetization factor N, of 0.1 – 0.592 (from N = $1/4\pi\chi_V + 1$), where $\chi_V$ = dM/dH is the value of linearly fitted slope for the up-right corner inset of **Figure 2d**. The experimental data can be fitted with the formula $M_{fit} = a + bH$ at low magnetic fields, where a is an intercept and b is a slope from fitting the low magnetic field magnetization measurements data. The bottom left corner inset of **Figure 2d** shows the M(H) − $M_{fit}$ data versus the magnetic field(H). $\mu_0H_{c1}^*$ was determined at the field when M deviates by ~ 1% above the fitted data ($M_{fit}$), as is the common practice.[35] We can obtain the lower critical field $\mu_0H_{c1}(T)$ in the consideration of the demagnetization factor (N), *via* the formula $\mu_0H_{c1}(T) = \mu_0H_{c1}^*(T)/ (1 - N)$. [36-37] The main panel of **Figure 2d** reveals the $\mu_0H_{c1}(T)$ as the function of temperature for $CuIr_2Te_4$. We estimated the $\mu_0H_{c1}(0)$ by fitting the $\mu_0H_{c1}(T)$ data *via* the formula $\mu_0H_{c1}(T) = \mu_0H_{c1}(0) [1 - (T/T_c)^2]$, which was shown by the black solid lines. The obtained zero-temperature lower critical field $\mu_0H_{c1}(0)$ for $CuIr_2Te_4$ was 0.028(2) T (**Table 2**).

With the purpose of estimating the critical field $\mu_0H_{c2}(0)$, we examined temperature dependent electrical resistivity at various applied fields $\rho(T, H)$ for $CuIr_2Te_4$ sample. **Figure 2e** exhibits the $\rho(T, H)$ measurement data for $CuIr_2Te_4$. **Figs. 2f** shows upper critical field values $\mu_0H_{c2}$ plotted *vs* temperature with $T_c$s obtained from resistivity at different applied fields. The $\mu_0H_{c2}$ *vs* $T$ curve near $T_c$ of $CuIr_2Te_4$ sample shows the well linearly fitting, which is represented by solid line. The value of fitting data slope ($dH_{c2}/dT$) of $CuIr_2Te_4$ sample was shown in **Table 2**. We can estimate the zero-temperature upper critical field (**Figure 2f**) of 0.12 T for $CuIr_2Te_4$ from the data, using the Werthamer-Helfand-Hohenberg (WHH) expression formula $\mu_0H_{c2} = -0.693T_c$ ($dH_{c2}/dT_c$) for the dirty limit superconductivity.[38-42] This obtained result was also summarized in **Table 2**. The Pauli limiting field of $CuIr_2Te_4$ was calculated from $\mu_0 H^P = 1.86 T_c$. The calculated values of $\mu_0 H^P$ was larger than the estimated values. Then, with this formula $\mu_0 H_{c2} = \frac{\phi_0}{2\pi\xi_{GL}^2}$, where $\phi_0$ is the flux quantum, the Ginzburg-Laudau coherence length ($\xi_{GL}(0)$) was calculated ~ 52.8 nm for $CuIr_2Te_4$ (**Table 2**).

For the sake of getting more abundant information of the electronic and superconducting properties of $CuIr_2Te_4$ polycrystalline sample, we took the heat capacity measurements for the polycrystalline sample under 0 T and 3 T applied magnetic field. The obtained $T_c$ from heat capacity matched the $T_c$ obtained by the $\chi(T)$ and $\rho(T)$ measurements well, which provides convincing evidence that a bulk superconducting state is achieved in $CuIr_2Te_4$. Thus, we have found a new layered quasi-two-dimensional ternary telluride $AB_2X_4$ chalcogenide superconductor wherein a superconducting state competes with CDW state, akin to the two-dimensional transition metal dichalcogenides (TMDCs). By fitting the normal state data of specific heats at high temperature under 3 T magnetic field, it is found to obey the formula $C_p = \gamma T + \beta T^3$, where $\gamma$ and $\beta$ describe the electronic and phonon contributions to the heat capacity, respectively.

We got the values of $\gamma$ and $\beta$ (inset of **Figure 5**) from fitting the data got under 3 T applied field in temperature range of 4 – 10 K. The normalized specific heat jump value $\Delta C/\gamma T_c$ obtained from the data (**Figure 3a**) was 1.82 for $CuIr_2Te_4$, which was much higher than the Bardeen-Cooper-Schrieffer (BCS) weak-coupling limit value (1.43), confirming bulk superconductivity.

Then we obtain the Debye temperature by the formula $\Theta_D = (12\pi^4 nR/5\beta)^{1/3}$ by using the fitted value of $\beta$, where $n$ is the number of atoms per formula unit and $R$ is the gas constant. Thus, we can estimate the electron-phonon coupling constant ($\lambda_{ep}$) by using the Debye temperature ($\Theta_D$) and critical temperature $T_c$ from the inverted McMillan formula: $\lambda_{ep} = \frac{1.04+\mu^* \ln\left(\frac{\Theta_D}{1.45T_c}\right)}{(1-1.62\mu^*)\ln\left(\frac{\Theta_D}{1.45T_c}\right)-1.04}$ [38]. This resultant $\lambda_{ep}$ is 0.65, suggesting that $CuIr_2Te_4$ belongs to an intermediately coupled superconductor. The electron density of states at the Fermi level ($N(E_F)$) can be calculated from $N(E_F) = \frac{3}{\pi^2 k_B^2 (1+\lambda_{ep})}\gamma$ with the $\gamma$ and $\lambda_{ep}$. This yield value that $N(E_F)$ = 2.72 states/eV f.u. for $CuIr_2Te_4$ (**Table 2**). The superconducting parameters of $CuIr_2Te_4$ are summarized in **Table 2**, and have a comparation with those sulpho- or seleno- spinels superconductors.

We now turn to the calculated electronic property of $CuIr_2Te_4$ in the disordered trigonal structure. The optimized structure of our model shows the interlayer distances between $IrTe_2$ layers are 5.21 Å and 5.60 Å, in the presence and absence of Cu intercalation, respectively. The former is close to that of 5.1 Å seen in **Figure1** (b). Compared with pure $IrTe_2$ with $c$ = 5.3984 Å,[43] the intercalation of Cu tends to reduce the interlayer distance (**Table 1**), and the interlayer Te-Te bonding [44] is expected to be stronger. This can be seen from the band structure, as shown in **Figure 4**, where the bands disperse strongly along ΓA. As revealed by both orbital projected band structure and density of state, the bands near the Fermi energy $E_F$ mainly come from Te $p$ and Ir $d$ orbitals, similar to that of $CuIr_2S_4$ in spinel structure.[46] Moreover, note that the density of states at the Fermi energy is 1.65, smaller than the experimentally derived value. Since the element ratio of Cu, Ir and Te is close to 1: 1.9: 4 in our sample, the 0.1 Ir vacancy can be thought of hole doping, thus the experimental value corresponds to the density of states at somewhere below the Fermi energy (~ 0.1 eV), which agrees better with the calculated result. Furthermore, one can notice that the Fermi energy locates at a flat plateau, not a local maximum, in the density of states. This behavior is also seen in $CuRh_2S_4$ [46], $CuV_2S_4$ [47] and monolayer $NbSe_2$ [48]. Although the structure of the CDW phase is still unknown, the change of density of states due to CDW transition is expected to be small,[43] and our comparison between the experimental and calculated values is reasonable.

**Figure 5** shows the Fermi surface, which consists of two parts, the outer Fermi surface is flower-like and the inner one is vase-like. Inside the inner Fermi surface, there is another egg-like Fermi surface surrounding Γ not seen in the figure. The shape of the two Fermi surfaces varies along ΓA, and remarkable nesting effects of the Fermi surface should not be expected.[46] Spin polarized calculation was further performed to confirm that no magnetism appears, and the rising behavior of magnetic susceptibility at low temperature (**Figure 2c**) was not from splitting of bands due to spin polarization under magnetic field. Compared with *Ref. 27*, where the magnetic susceptibility satisfies the Curie-Weiss law, our data here is more complicated in that the susceptibility remains nearly constant in the temperature range of 50 – 200 K, and can't be described solely by the Pauli paramagnetism. Since no magnetism appears, the spin fluctuation due to the divergence of spin susceptibility

$$\chi^s(q, i\omega_n)=[1-\chi^0(q, i\omega_n)U]^{-1}\chi^0(q, i\omega_n)$$

does not play a role, where $\chi^s(q, i\omega_n)$, $\chi^0(q, i\omega_n)$ are the spin susceptibility with and without correlation $U$, respectively. Therefore, the superconducting transition temperature should be low, as what is found here.[49] To see the origin of the CDW phase we calculated the phonon dispersion of $CuIr_2Te_4$ (**Figure 6**). As can be seen, neither imaginary branches nor unusual discontinuity dip can be found, suggesting that the CDW is not driven by Kohn anomaly [45] similar to pure $IrTe_2$.[43] Since the wave vector characterizing the CDW phase is not determined in our work, the superlattice corresponding to CDW is thus unknown.

However, the resistivity anomaly near around 250 K shows CDW character. With the evidence we have, the sudden rise in resistivity at 250 K, as temperature is lowered, is expected to come from the reduced density of states at Fermi energy. We suspect this reduced density of state is due to a sudden increase in the correlation between Ir d electrons during their hopping process, as temperature is lowered. At low temperature the frequency disperses linearly with the wave vector, and the heat capacity due to phonon part var-ies linearly with T3. We deduced the coefficient to be 1.77, as shown in Figure 3b. The deduced Debye temperature is 197.36 K, agrees well with the experimentally derived value.

## CONCLUSION

Here we report the first observation of the coexistence of the superconductivity and charge density wave in the $AB_2X_4$-type ternary telluride $CuIr_2Te_4$, which was synthesized *via* a solid-state reaction method. From the results of magnetization, resistivity and heat capacity measurements we find that $CuIr_2Te_4$ is a quasi-two-dimensional ternary telluride chalcogenide superconductor with a $T_c \approx 2.5$ K and coexists a CDW state around 250 K. Electronic structure calculation shows the states near the Fermi energy mainly come from Te *p* and Ir *d* orbitals. The lack of nesting in the Fermi surface and the missing of unusual discontinuity in phonon spectrum exclude the possibility that the CDW phase is driven by Fermi surface nesting and Kohn anomaly, respectively. Moreover, the calculated density of states at Fermi energy, phonon part of heat capacity at low temperature, and Debye temperature agrees semi-quantitatively with the experimental values, confirming the validity of our model. From more profound experimental and theoretical study CDW-bearing superconductor like $CuIr_2Te_4$, we can clarify the interplay between CDW and superconductivity quantum state hopefully in the condensed matter field.

**ACKNOWLEDGMENT**

The authors thank R. Cava, B. Shen and W.W. Xie for valuable discussions. H. X. Luo acknowledges the financial support by "Hundred Talents Program" of the Sun Yat-Sen University and Natural Science Foundation of China (21701197). J. Ma is supported by the National Natural Science Foundation of China (Grant No. 11774223). Y. Zeng and D. X. Yao are supported by NKRDPC Grants No. 2017YFA0206203, No. 2018YFA0306001, NSFC-11574404, National Supercomputer Center in Guangzhou, and Leading Talent Program of Guangdong Special Projects. T.-R.C. was supported by the Young Scholar Fellowship Program from the Ministry of Science and Technology (MOST) in Taiwan, under a MOST grant for the Columbus Program MOST108-2636-M-006-002, National Cheng Kung University, Taiwan, and National Center for Theoretical Sciences, Taiwan. This work was supported partially by the MOST, Taiwan, Grant MOST107-2627-E-006-001. This research was supported in part by Higher Education Sprout Project, Ministry of Education to the Headquarters of University Advancement at National Cheng Kung University (NCKU).

.

**Table 1.** Rietveld refinement structural parameters of $CuIr_2Te_4$. ($a$ = 3.9397(3) Å and $c$ = 5.3964(6) Å)

| Label | x | y | z | Site | OCC. |
|---|---|---|---|---|---|
| Ir | 0.00000 | 0.00000 | 0.00000 | 1a | 1.000 |
| Te | 0.33330 | 0.66670 | 0.2308(4) | 2d | 1.000 |
| Cu | 0.00000 | 0.00000 | 0.50000 | 1b | 0.500 |

**Table 2.** Comparison of superconducting parameters in $AB_2X_4$ superconductors

| Material | $CuIr_2Te_4$ | $CuRh_2S_4$ | $CuRh_2Se_4$ | $Cu_{0.7}Zn_{0.3}Ir_2S_4$ | $CuIr_{1.6}Pt_{0.4}Se_4$ | $CuV_2S_4$ |
|---|---|---|---|---|---|---|
| $T_c$ (K) | 2.50 | 4.7 | 3.5 | 3.4 | 1.76 | 4.45 |
| $\gamma_{exp}$ (mJ mol$^{-1}$ K$^{-2}$) | 10.57 | 26.9 | 21.4 | | 16.5 | |
| $ß_{exp.}$ (mJ mol$^{-1}$ K$^{-4}$) | 2.15 | | | | 1.41 | |
| $\Theta_D$ (K) | 185.5 | 258 | 218 | | 212 | |
| $\Delta C/\gamma T_c$ | 1.82 | 1.89 | 1.68 | | 1.58 | |
| $\lambda_{ep(exp.)}$ | 0.65 | 0.66 | 0.63 | | 0.57 | |
| $N(E_F)_{exp.}$(states/eV f.u) | 2.72 | | | | 4.45 | |
| $-dH_{c2}/dT$ (T/K) | 0.066 | 0.614 | 0.181 | | 2.62 | |
| $\mu_0H_{c2}$(T) | 0.12 | 2.0 | 0.44 | | 3.2 | |
| $\mu_0H^P$(T) | 4.65 | 8.74 | 6.51 | 6.32 | 3.27 | |
| $\mu_0H_{c1}$(T) | 0.028 | | | | | |
| $\xi_{GL}(0)$ (nm) | 52.8 | | | - | 0.96 | |
| $\gamma_{cal.}$ (mJ mol$^{-1}$ K$^{-2}$) | 6.42 | | | | | |
| $ß_{cal.}$ (mJ mol$^{-1}$ K$^{-4}$) | 1.77 | | | | | |
| $\lambda_{ep(cal.)}$ | | | | | | |
| $N(E_F)_{cal.}$(states/eV f.u) | 1.65 | | | | | |
| $\Theta_{D,cal}$ (K) | 197.36 | | | | | |

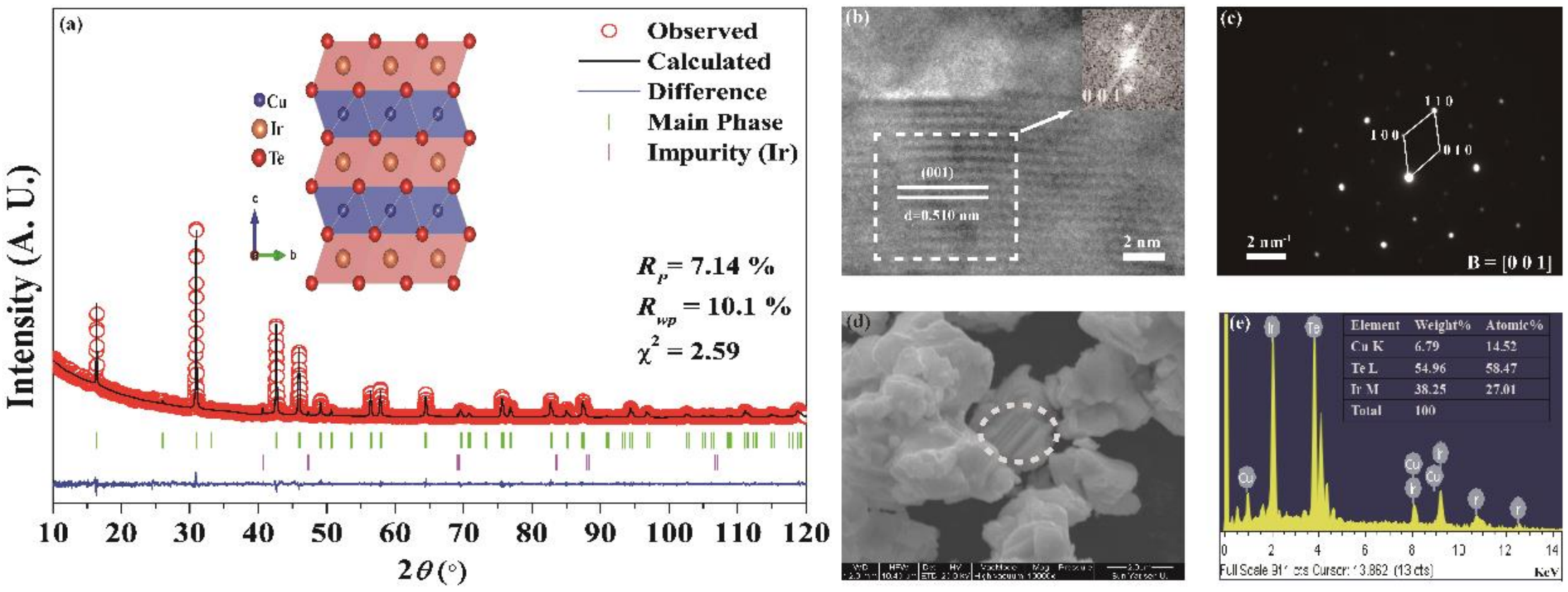

**Figure 1**. (a) Powder XRD pattern with Rietveld refinement for $CuIr_2Te_4$. Inset of (a) shows the view of $CuIr_2Te_4$ structure along (100) direction. (b) HRTEM image of $CuIr_2Te_4$ polycrystalline powder. Inset is the corresponding FFT pattern. (c) Selected area (100) direction. (b) HRTEM image of $CuIr_2Te_4$ polycrystalline powder. Inset is the corresponding FFT pattern. (c) Selected area electron diffraction pattern of $CuIr_2Te_4$ polycrystalline powder taken on [001] orientation. (d) SEM image of $CuIr_2Te_4$ polycrystalline powder in the magnification of 13000 times. (e) EDS spectrum and element ratio of $CuIr_2Te_4$.

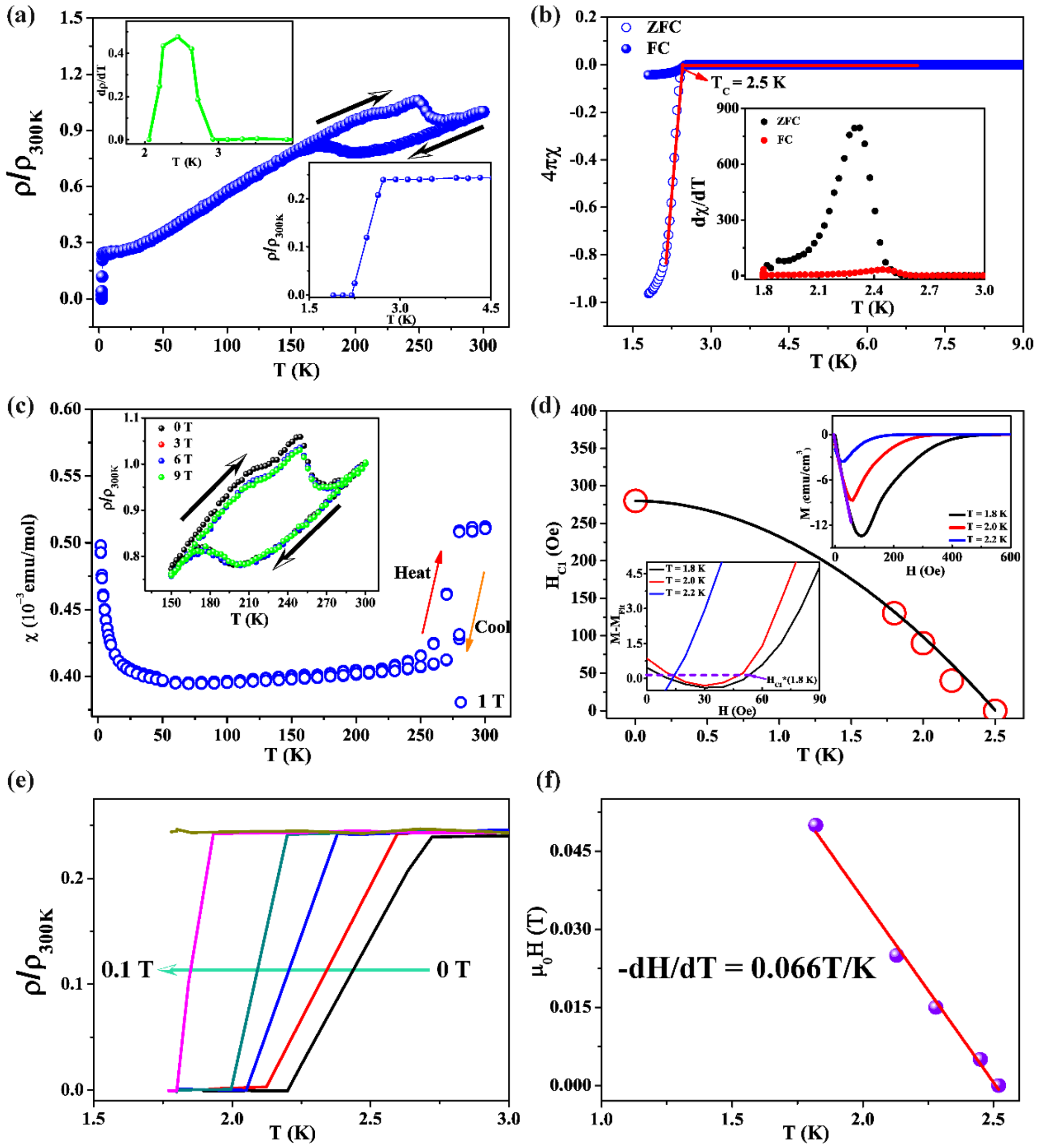

**Figure 2**. Transport characterization of the normal states and superconducting transitions for $CuIr_2Te_4$. (a) The temperature depend-ence electrical resistivity for polycrystalline $CuIr_2Te_4$. Bottom right corner inset of (a) shows the temperature dependence electrical resistivity at low temperature. Up left corner inset of (a) shows metallic temperature dependence ($d\rho/dT$) in the temperature region of 2 - 5 K. (b) Magnetic susceptibility for $CuIr_2Te_4$ at the superconducting transitions; applied DC field was 20 Oe. Both zero-field cooling (ZFC) and field cooling (FC, shown more clearly in the inset) curves were measured. Inset of (b) shows the $d\chi/dT$ vs T curve at low temperature. (c) Magnetic susceptibility of $CuIr_2Te_4$ as a function of temperature at applied field of 1 T. Inset of (c) shows the CDW state at applied fields. (d) Temperature dependence of the lower critical field ($\mu_0 H_{c1}$) for $CuIr_2Te_4$. Up right corner inset shows magnetic susceptibility at low applied magnetic fields at various applied temperatures for $CuIr_2Te_4$. Bottom left inset shows $M$-$M_{fit}$ vs H. (e) Low temperature resistivity at various applied fields for $CuIr_2Te_4$. (f) $\mu_0 H(T)$ at different $T_c$s, red solid line shows linearly fitting to the data to estimate $\mu 0 H_{c2}$.

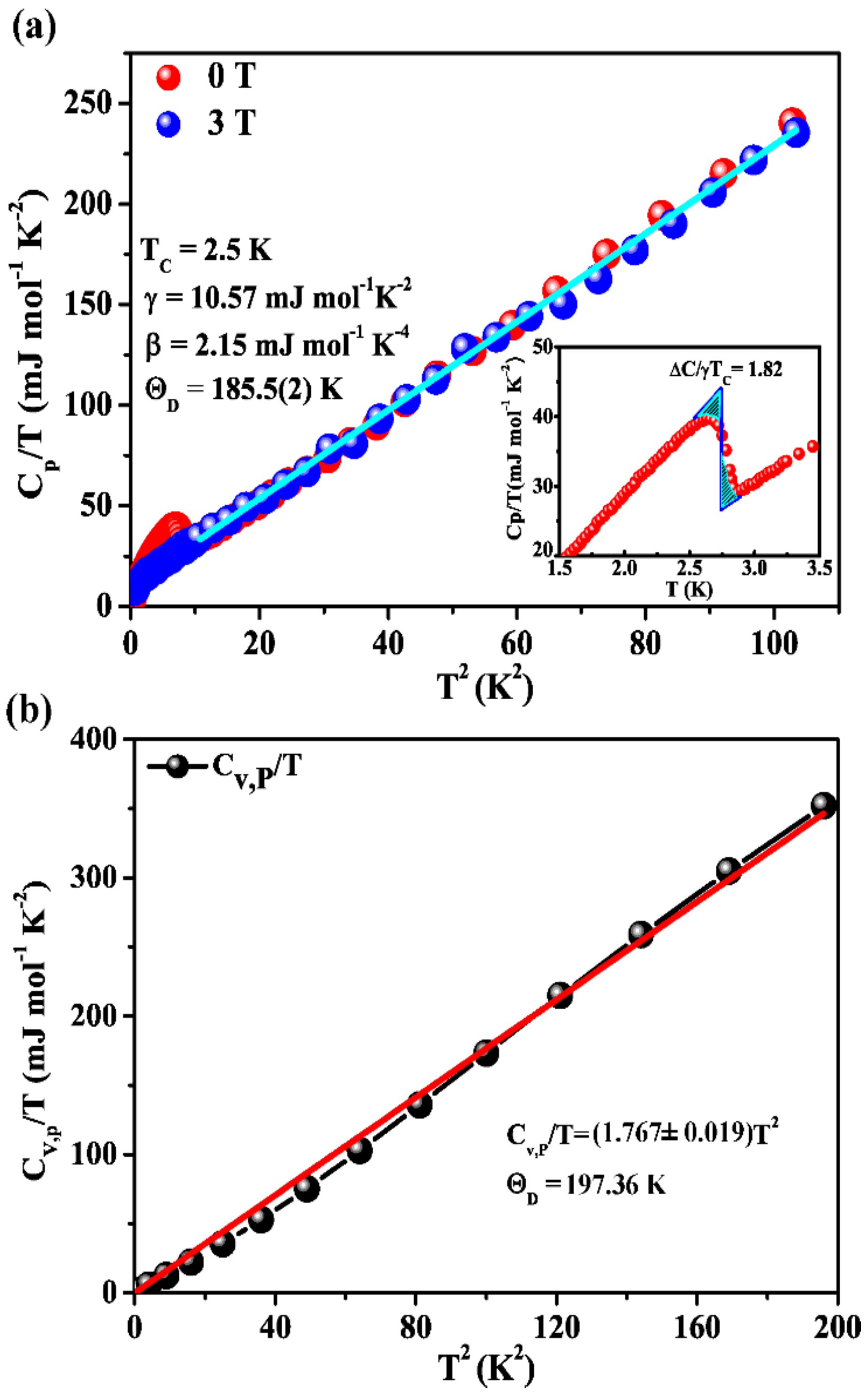

**Figure 3**. (a) Heat Capacity characterization of $CuIr_2Te_4$. Debye temperature of $CuIr_2Te_4$ obtained from fits to data in applied field of 3 T. Inset show heat capacities through the superconducting transitions without applied magnetic field for $CuIr_2Te_4$. (b) The heat capacity derived from the phonon contribution at low temperature for $CuIr_2Te_4$.

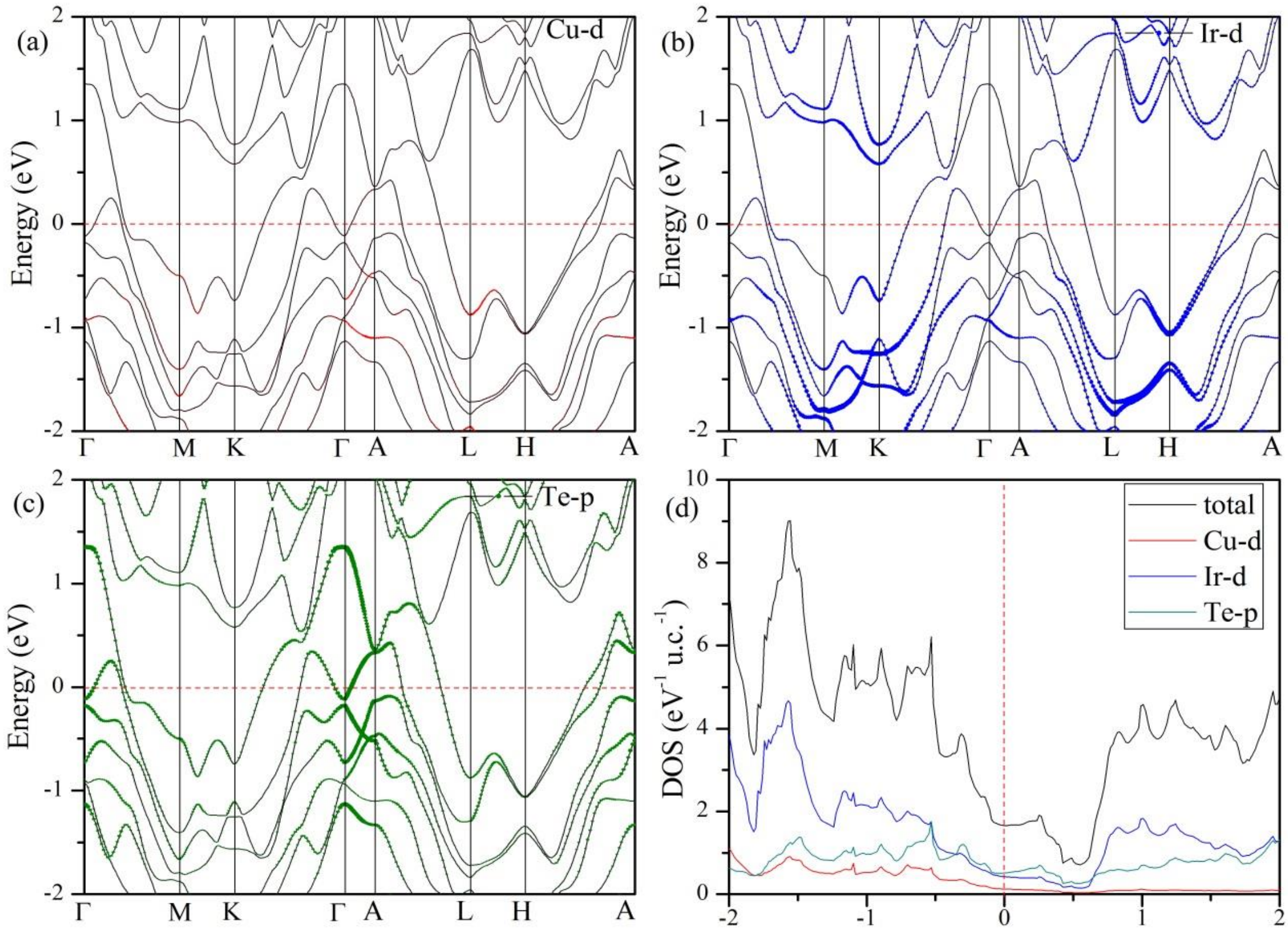

**Figure 4**. The orbital-projected band structure of $CuIr_2Te_4$ with SOC projected at (a) Cu d, (b) Ir d and (c) Te p orbitals. The orbital weight at energy eigenvalue $E_{nk}$ is shown by the size of the points. (d) is the total and partial density of states near the Fermi energy.

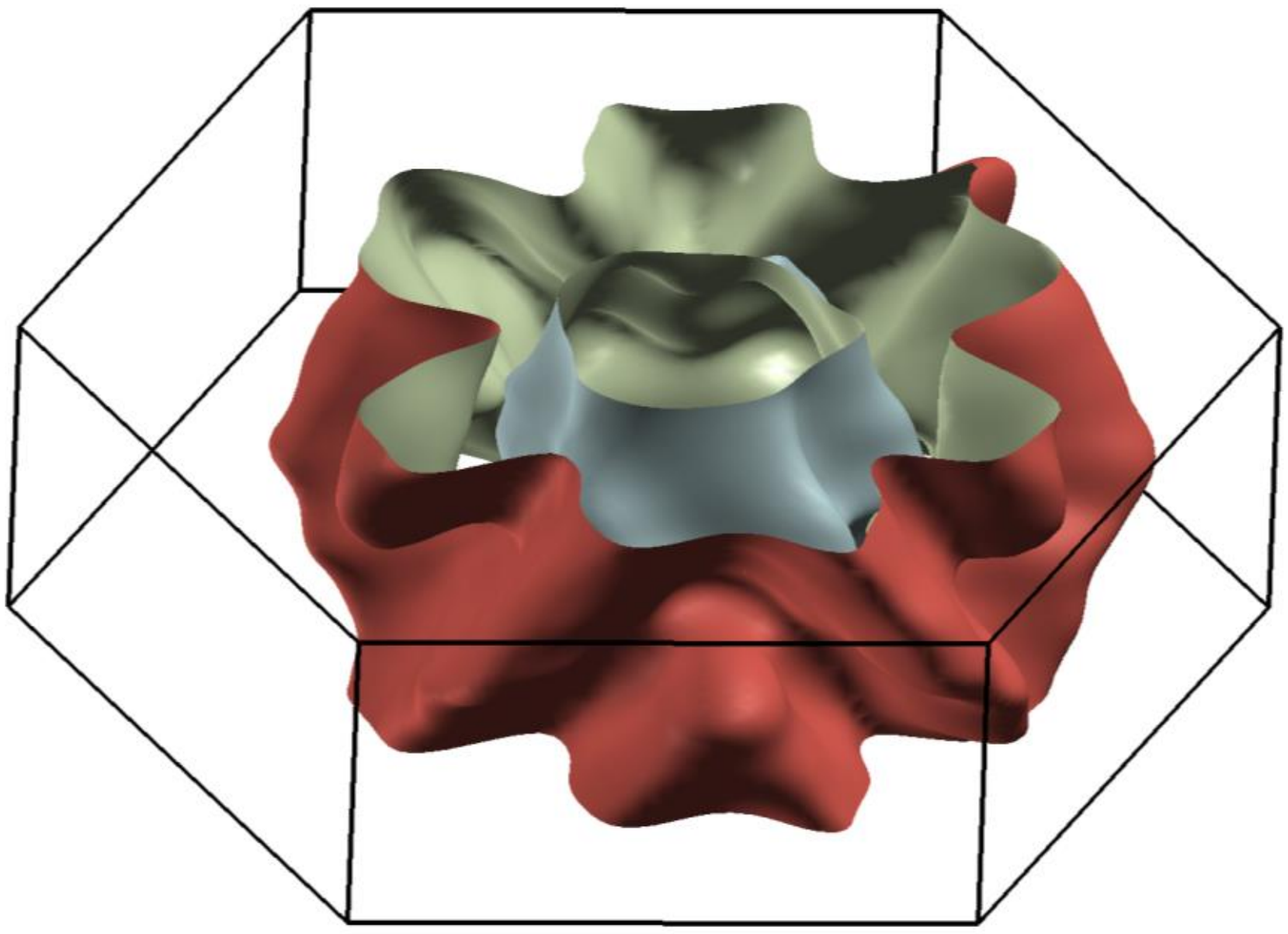

**Figure 5.** The Fermi surface of $CuIr_2Te_4$ with SOC.

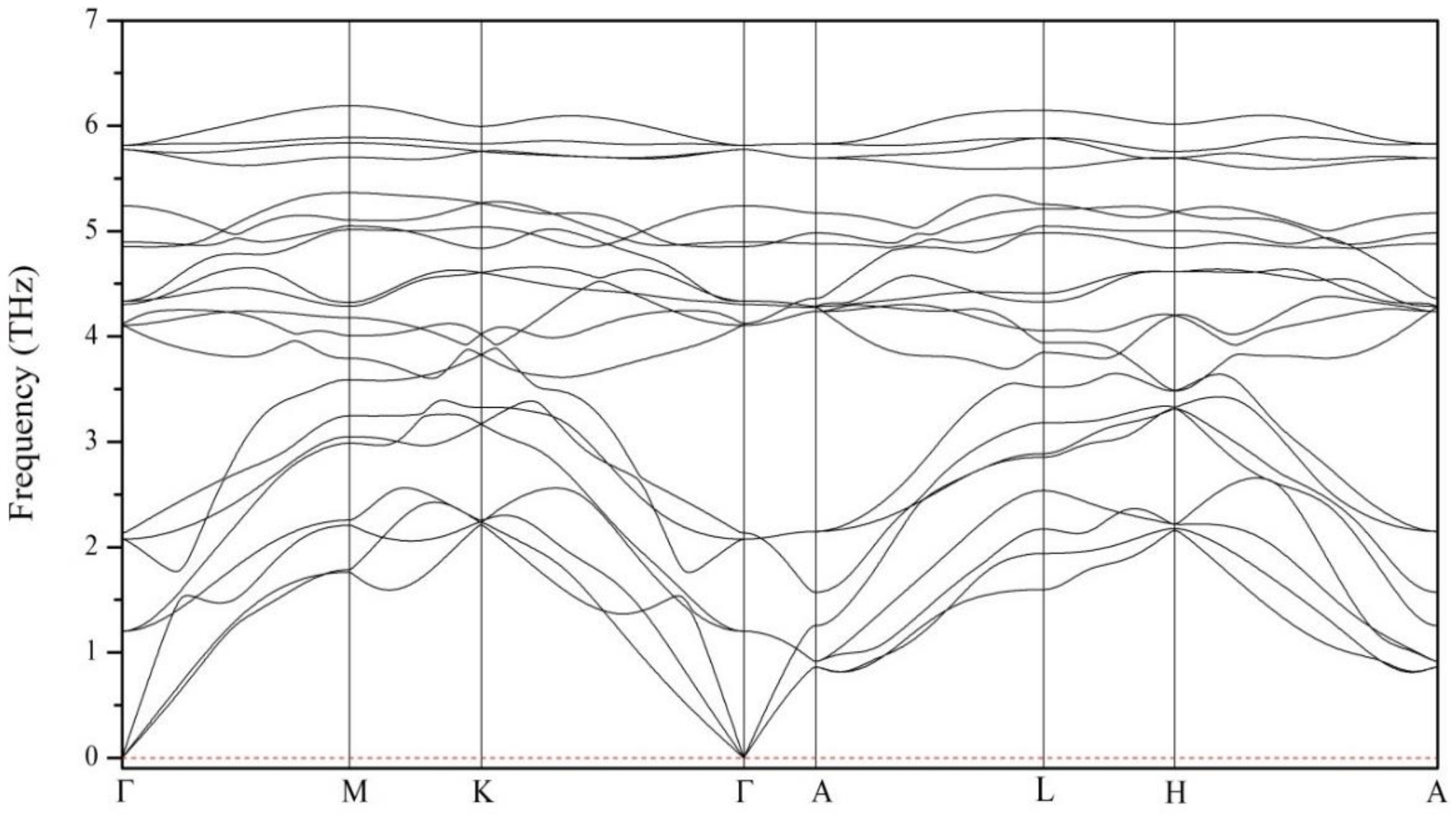

**Figure 6.** The calculated phonon dispersion of $CuIr_2Te_4$.